\documentclass[paper]{elsarticle}

\usepackage{amsfonts,amsbsy,amssymb,amsmath}
\usepackage{hyperref}
\usepackage{graphicx}
\usepackage[table,xcdraw]{xcolor}
\usepackage{float}

\journal{Chemical Physics Letters}

\biboptions{compress}

\begin{document}

\begin{frontmatter}

\title{The Dynamical Matching Mechanism in Phase Space for 
Caldera-Type Potential Energy Surfaces}

\author[label1]{M. Katsanikas}
\author[label2]{V. J. Garc\'ia-Garrido}
\author[label1]{S. Wiggins\corref{mycorrespondingauthor}}
\ead{S.Wiggins@bristol.ac.uk}

\address[label1]{School of Mathematics, University of Bristol, \\ Fry Building, Woodland Road, Bristol, BS8 1UG, United Kingdom.\\[.2cm]}

\address[label2]{Departamento de F\'isica y Matem\'aticas, Universidad de Alcal\'a, \\ Alcal\'a de Henares, 28871, Spain.}

\cortext[mycorrespondingauthor]{Corresponding authors}

\begin{abstract}
Dynamical matching occurs in a variety of important organic chemical reactions. It is observed to be a result of a potential energy surface (PES) having specific geometric features. In  particular, a region of relative flatness where entrance and exit to this region is controlled by index-one saddles. Examples of potential energy surfaces having these features are the so-called caldera potential energy surfaces. We develop a  predictive level of understanding of the phenomenon of dynamical matching in a caldera potential energy surface. We show that the phase space structure that governs dynamical matching is a particular type of heteroclinic trajectory which gives rise to trapping of trajectories in the central region of the caldera PES. When the heteroclinic trajectory is broken, as a result of parameter variations, then dynamical matching occurs.
\end{abstract}


\end{frontmatter}


\section{Introduction}
Dynamical matching is an interesting chemical dynamical phenomenon that occurs in a variety of  organic chemical reactions. For example, it occurs in the vinylcyclopropane-cyclopentene rearrangement \cite{baldwin2003,gold1988}, the stereomutation of cyclopropane \cite{doubleday1997}, the degenerate rearrangement of bicyclo[3.1.0]hex-2-ene \cite{doubleday1999,doubleday2006} or that of 5-methylenebicyclo[2.1.0]pentane \cite{reyes2002}. A general description of the phenomenon was first given by Carpenter in \cite{carpenter1985,carpenter1995} where it was argued that a general type of potential energy surface exhibiting the dynamical matching phenomenon has the landscape  resembling that of a caldera.

The potential energy landscape of a Caldera gets its name from the shape of the region corresponding to an erupted volcano. It is characterized by a shallow, almost ‘’flat’’  potential well region (a central minimum), surrounded by four entrance/exit channels in the form of index-1 saddles. Two of these saddles have low energy values and correspond to the formation of chemical products, while the other two are higher in energy and represent reactants. 

The manifestation of the dynamical matching phenomenon is essentially a statement of  momentum conservation and Newton’s first law of motion. It is observed that a trajectory entering the Caldera from a channel corresponding to a high energy index-1 saddle (reactant) experiences little force in the caldera due to the ‘’flatness’’ of the PES, and it exits through the diametrically opposing low energy index-1 saddle (product). Consequently, this mechanism determines to a considerable extent the outcome of the chemical reaction. However, not all trajectories entering the caldera experience dynamical matching. It is observed that some trajectories may interact with the shallow potential well and become temporarily trapped in the region. This can dramatically influence the manner in which they exit from the well. 

A detailed study of the trajectory behavior in a two degree-of-freedom (DoF) caldera PES was given in \cite{collins2014}, where a more general discussion of caldera-like PESs in organic reactions is also presented. Further work elucidating the phenomena of dynamical matching and trapping  in this caldera model was carried out in \cite{ katsanikas2018,katsanikas2019}. We will describe the results in these papers in more detail when we describe the Hamiltonian model in the next section.

In this paper we describe the phase space mechanism that controls dynamical matching. We show that  dynamical matching is controlled by a heteroclinic intersection between the unstable manifold of a periodic orbit controlling entrance to the caldera and the stable manifold of a periodic orbit  in the region of the shallow minimum.  When a heteroclinic connection exists, trajectories thst enter the caldera are transported to the shallow minimum, and they experience temporary trapping in this region. When there is no heteroclinic connection, trajectories enter and exit the caldera without interacting with the  region of the central minimum.  Knowledge of this phase space mechanism is significant because it allows us to predict existence, and non-existence, of dynamical matching.

This paper is outlined as follows. In Section \ref{sec:model} we describe the Caldera PES that we use in this work, its critical points and stability, and the resulting Hamiltonian model. Section \ref{DM_mech} is devoted to analyzing the phase space structures that govern dynamical matching. Finally, in Section \ref{CONCL} we summarize the results and give an outlook for future research directions.

\section{The Hamiltonian Model of the Caldera}
\label{sec:model}
 
\begin{figure}[htbp]
\begin{center}
\includegraphics[scale=0.38]{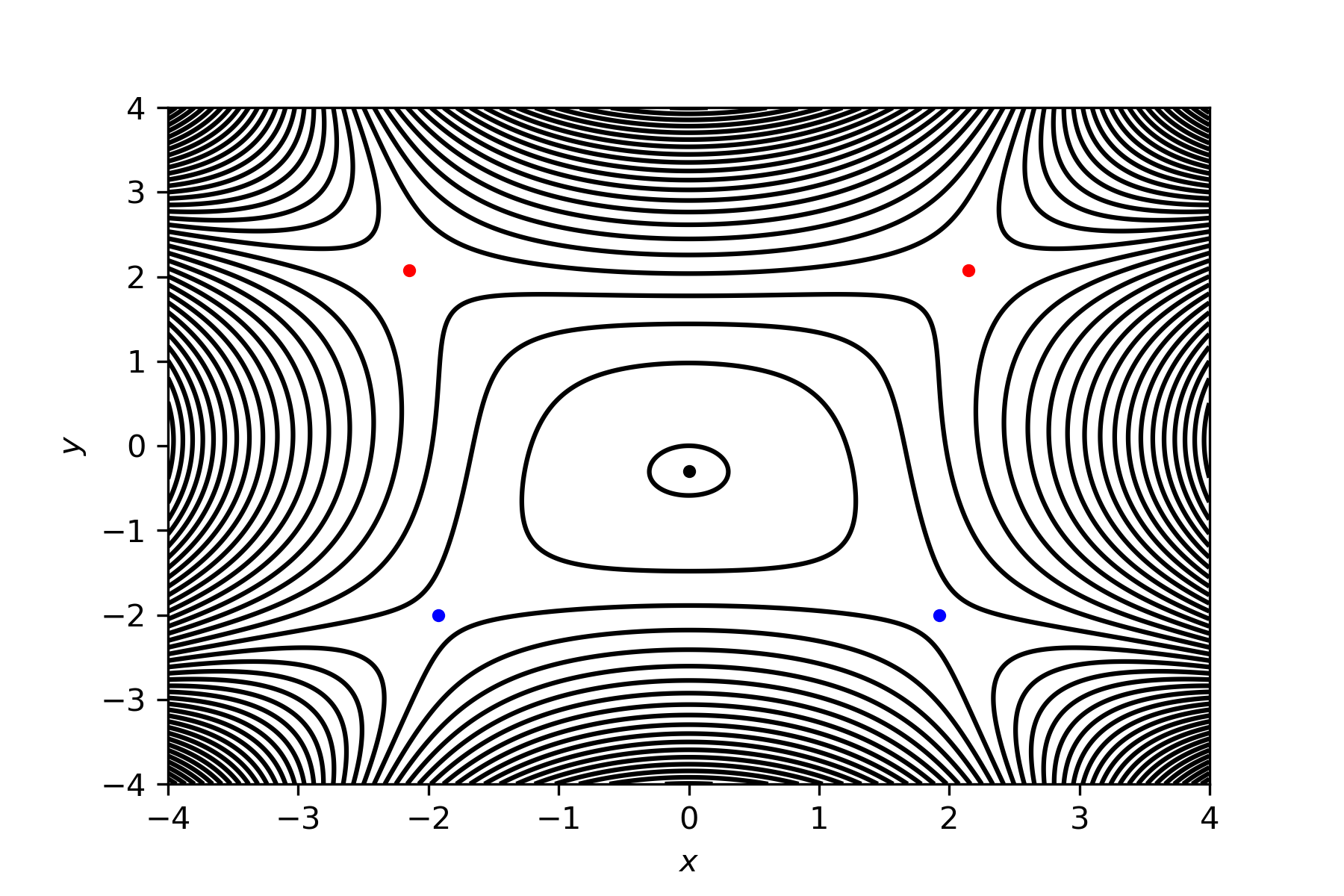}
\includegraphics[scale=0.38]{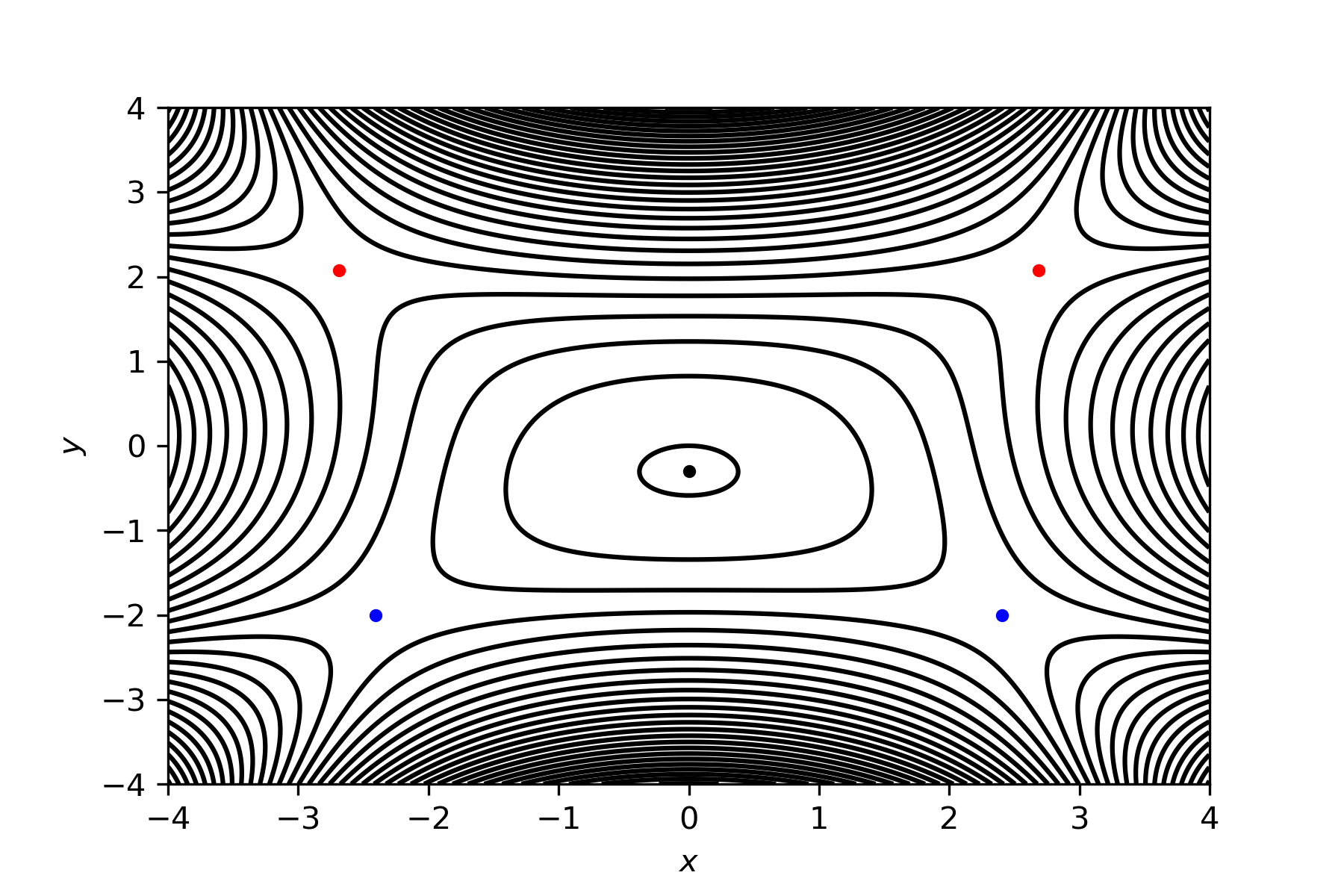}\\
\includegraphics[scale=0.38]{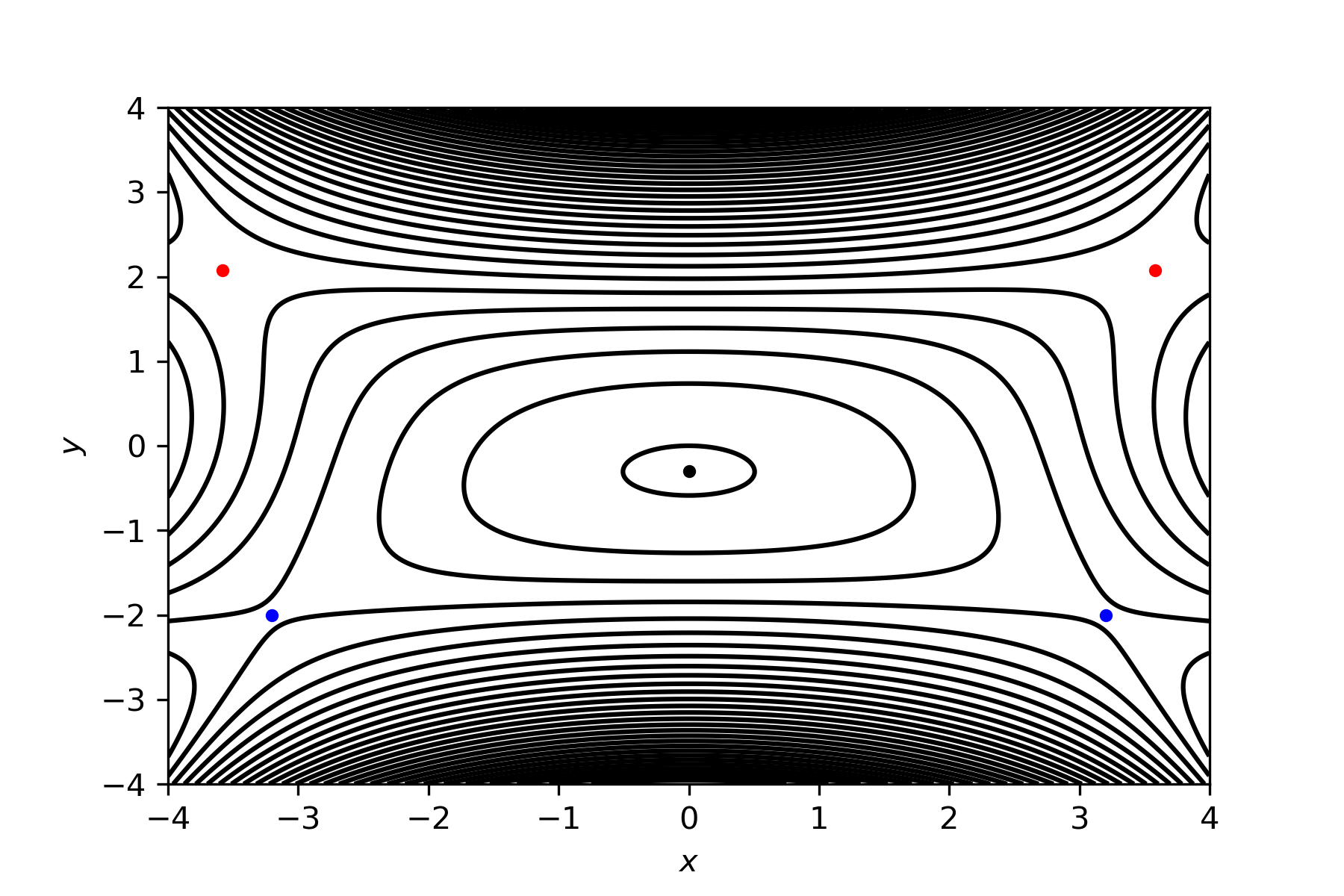}
\includegraphics[scale=0.38]{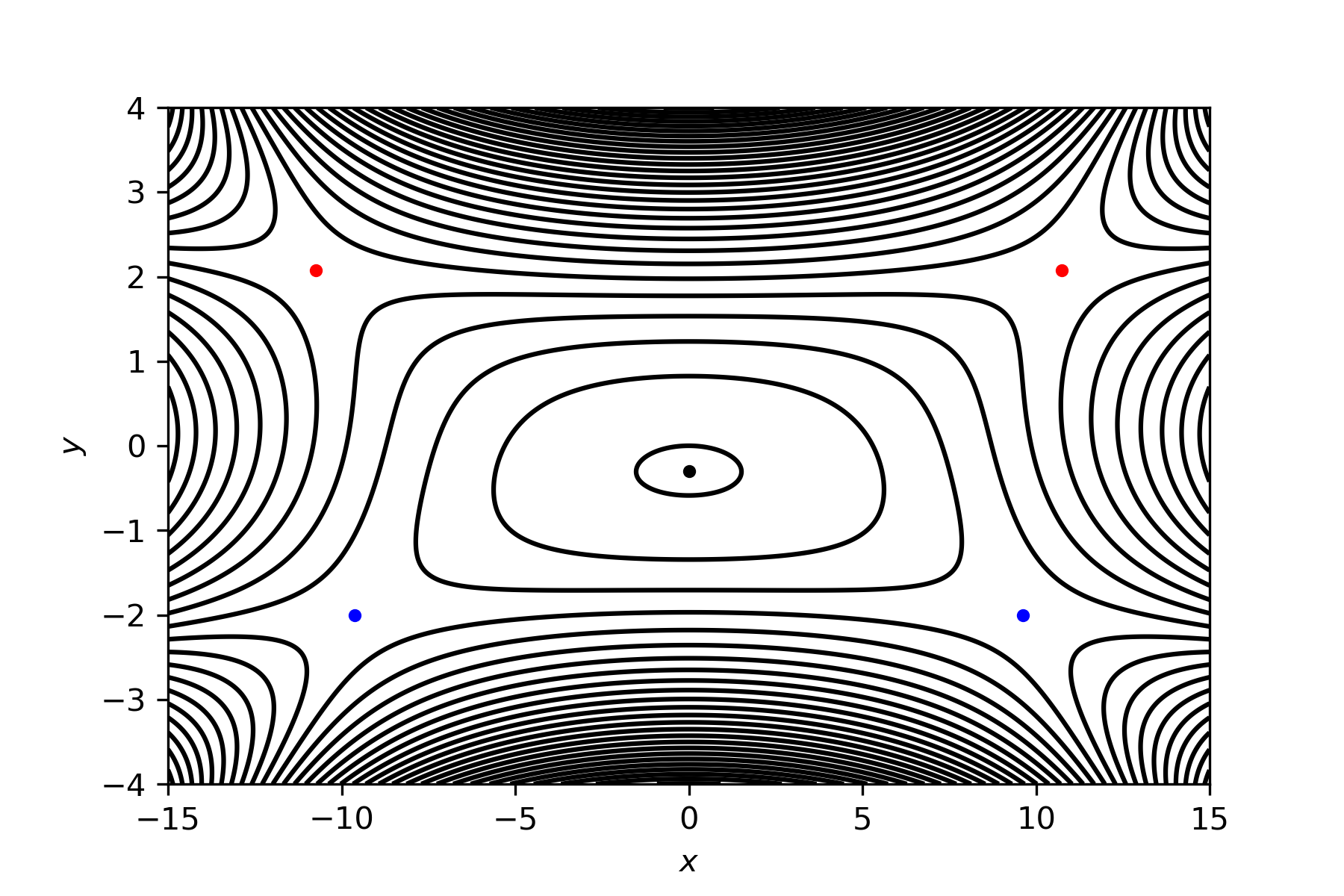}\\
\end{center}
\caption{The stable equilibrium point in the center (depicted by a black point), the upper saddles (depicted by red points), the lower saddles (depicted by blue points) and the  contours of the potential for $\lambda = 1$ (upper left panel), $\lambda = 0.8$ (upper right panel), $\lambda = 0.6$ (lower left panel) and  $\lambda = 0.2$ (lower right panel).}
\label{equi}
\end{figure}

In this section we describe the Caldera potential energy surface that we consider. This model has previously been studied in \cite{collins2014,katsanikas2018,katsanikas2019}.
The key features of the PES are a central minimum surrounded by four index-1 saddles. Two of the saddles are of higher energy than the others and phase space structures associated with these saddles control entrance and exit into the caldera. The PES, which contains a parameter $\lambda$ that controls the stretching of the potential in the horizontal direction and breaks the symmetry is given by:
\begin{equation}
V(x,y) = c_1 \left(y^2 + (\lambda x)^2\right) + c_2 \, y - c_3 \left((\lambda x)^4 + y^4 - 6 \, (\lambda x)^2 y^2\right)
\label{eq1}
\end{equation}
\noindent
The parameters used in this paper are $c_1 = 5$, $c_2 = 3$, $c_3 = -3/10$ and $0 < \lambda \leq 1$ (the stretching parameter). The  symmetric caldera PES \cite{collins2014,katsanikas2018} corresponds to $\lambda = 1$ and is shown in the upper left hand panel of  Fig. \ref{equi}. In the remaining panels of  Fig. \ref{equi} we depict the PES contours and the equilibrium points for $\lambda=0.8$, $\lambda=0.6$ and $\lambda=0.2$. Table \ref{tab:ta08} gives the positions and energies of the upper index-1 saddles for the different values of $\lambda$ shown in Fig. \ref{equi}. We observe that the positions of the index-1 saddles move away from the center of the Caldera as we decrease the parameter $\lambda$, which is a consequence of stretching the PES. The position of the central minimum is given by $(x,y) = (0,-0.297)$ with energy $E = -0.448$ for all values of the stretching parameter $\lambda$.

\begin{table}[htbp]   
	\begin{center} 
   \begin{tabular}{| l c c c |}
          \hline
          Equilibrium point & x & y & $\lambda$ \\
          \hline
          Upper LH index-1 saddle  &-2.149   & 2.0778 & 1 \\
          Upper RH index-1 saddle  &2.149    &  2.0778 & 1\\
          Upper LH index-1 saddle  &-2.6862  & 2.0778 & 0.8 \\
          Upper RH index-1 saddle  &2.6862  &  2.0778 & 0.8\\
          Upper LH index-1 saddle  &-3.5815  & 2.0778 & 0.6 \\
          Upper RH index-1 saddle  &3.5815  &  2.0778 & 0.6 \\
          Upper LH index-1 saddle  &-10.7446 & 2.0778 & 0.2 \\
          Upper RH index-1 saddle  &10.7446 &  2.0778 & 0.2 \\
          \hline
 \end{tabular} 
\end{center} \label{tab:ta08}
\caption{The upper index-1 saddles of the PES in Eq. \ref{eq1} ("RH" and "LH" are the abbreviations for right hand and left hand respectively) for different values of $\lambda$. The energy for each of the cases is $E = 27.0123$.} 
\end{table}
 
The Hamiltonian for the system with two DoF is the sum of kinetic plus potential energy:
\begin{equation}
H(x,y,p_x,p_y) = \frac{p_x^2}{2m_x} + \frac{p_y^2}{2m_y} + V(x,y)
\label{eq2}
\end{equation}
\noindent
where $V(x,y)$ is the Caldera PES in Eq. \eqref{eq1}, and $m_x$, $m_y$ are the masses of the $x$ and $y$ DoF respectively. In this work, for simplicity, we take $m_x = m_y = 1$. Hamilton's equations of motion are given by:
\begin{equation}
\begin{cases}
\dot x = \dfrac{\partial H} {\partial p_x} = \dfrac{p_x}{m_x} \\[.4cm]
\dot y = \dfrac{\partial H} {\partial p_y} = \dfrac{p_y}{m_y} \\[.4cm]
\dot p_x = -\dfrac{\partial H} {\partial x} = 2 \lambda \, (\lambda x) \left[2c_3 \left((\lambda x)^2 - 3 y^2 \right) - c_1 \right] \\[.4cm]
\dot p_y = -\dfrac {\partial H} {\partial y} = 2 y \left[ 2 c_3 \left(y^2 - 3 (\lambda x)^2\right) - c_1 \right] - c_2
\end{cases}
\label{eq3}
\end{equation}

\section{The Phase Space Mechanism Governing Dynamical Matching}
\label{DM_mech}

As we have described in the introduction, the caldera gets its name from the shape of the PES. However, transport across the caldera is a dynamical phenomenon governed by the template of geometrical structures in phase space, and dynamical matching is just one particular type of  dynamical phenomenon that we are considering in this paper. First, we describe the phase space structures that mediate transport into the caldera.

In order to reveal the phase space structures that are responsible for the mechanism that allows and prevents dynamical matching, we use in this work the method of Lagrangian descriptors (LDs), see e.g. \cite{mancho2013lagrangian,lopesino2017,naik2019a}. Lagrangian descriptors is a trajectory-based scalar diagnostic that has been developed in the nonlinear dynamics literature to explore the geometrical template of phase space structures that characterizes qualitatively distinct dynamical behavior. The technique was originally developed for studies of transport and mixing in geophysical flows \cite{madrid2009} but has recently been applied to problems in chemical reaction dynamics  e.g \cite{craven2015lagrangian,craven2016deconstructing,craven2017lagrangian}, where the computation of chemical reaction rates relies on the knowledge of the phase space structures that separate reactants from products. Recent modifications of this technique, known as variable integration time Lagrangian descriptors, have made it applicable to finding phase space structure in open Hamiltonian systems \cite{junginger2017chemical,naik2019b,GG2019}. Details on how they are applied for revealing phase space structures in caldera-like PESs are described in \cite{KGW2019}. In this short paper we focus on presenting the results relevant to dynamical matching.

\begin{figure}[htbp]
	\centering
	\includegraphics[scale=0.25]{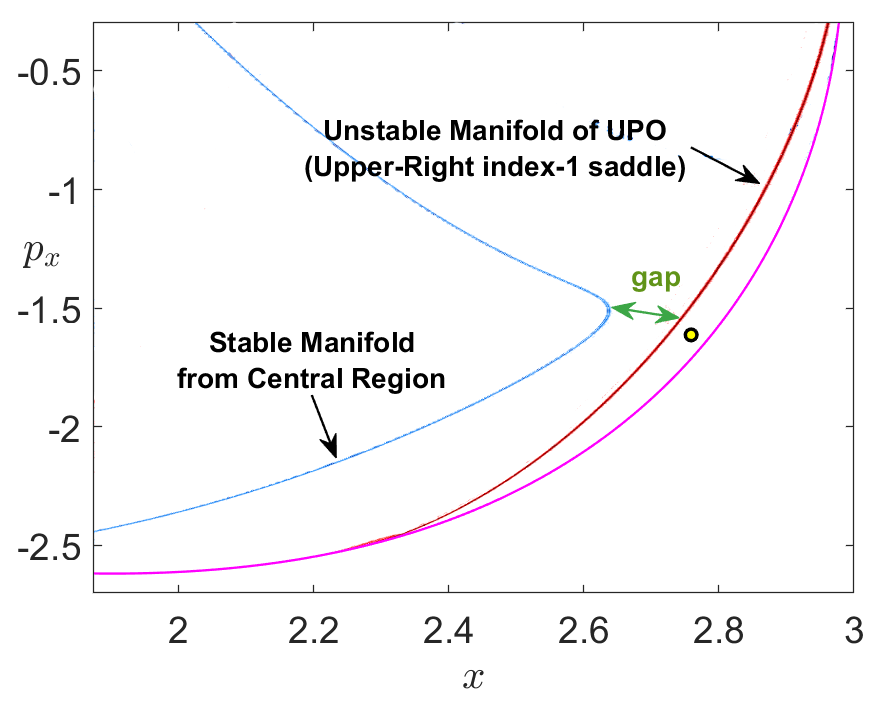}
	\includegraphics[scale=0.29]{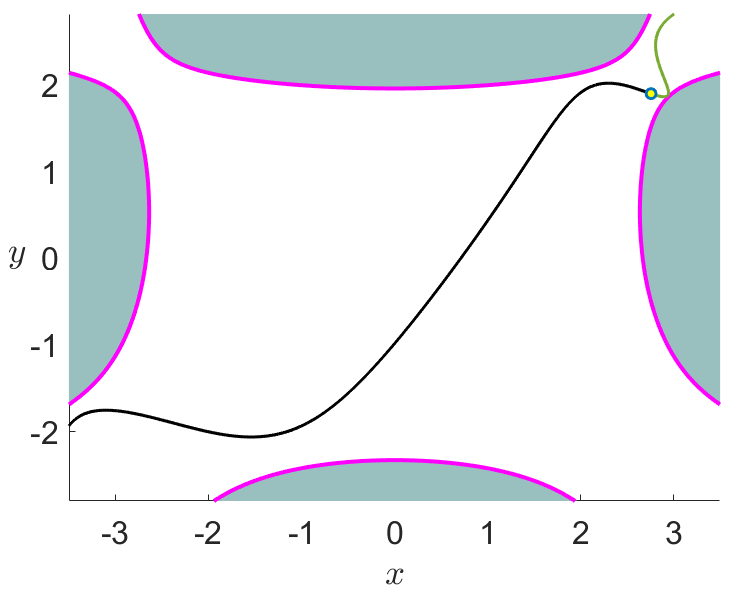}\\
	\includegraphics[scale=0.25]{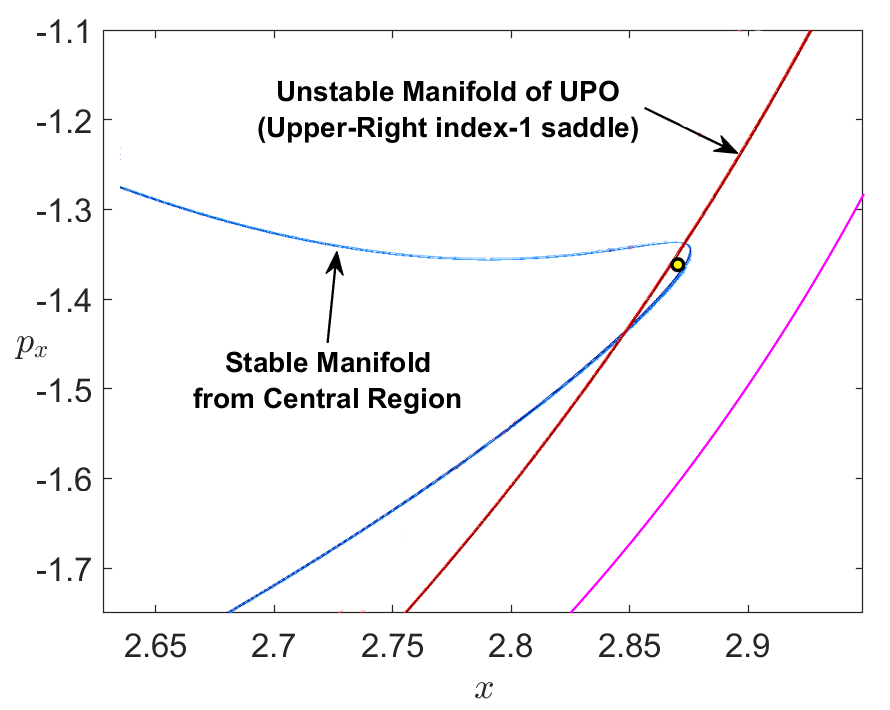}
	\includegraphics[scale=0.29]{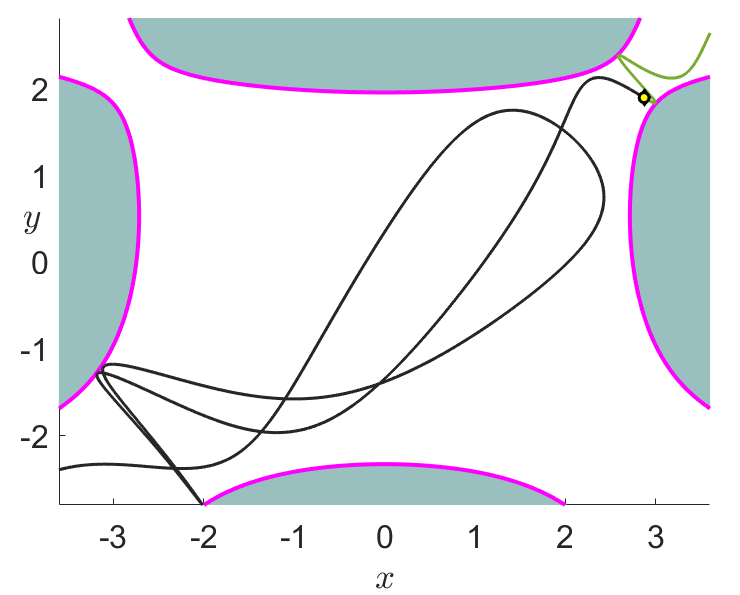}\\
	\includegraphics[scale=0.25]{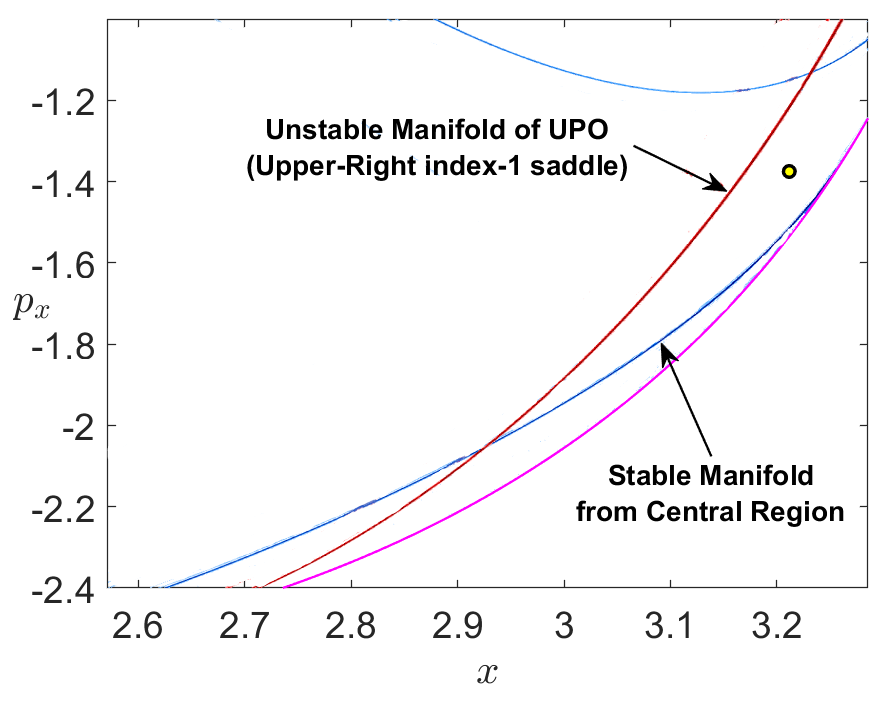}
	\includegraphics[scale=0.29]{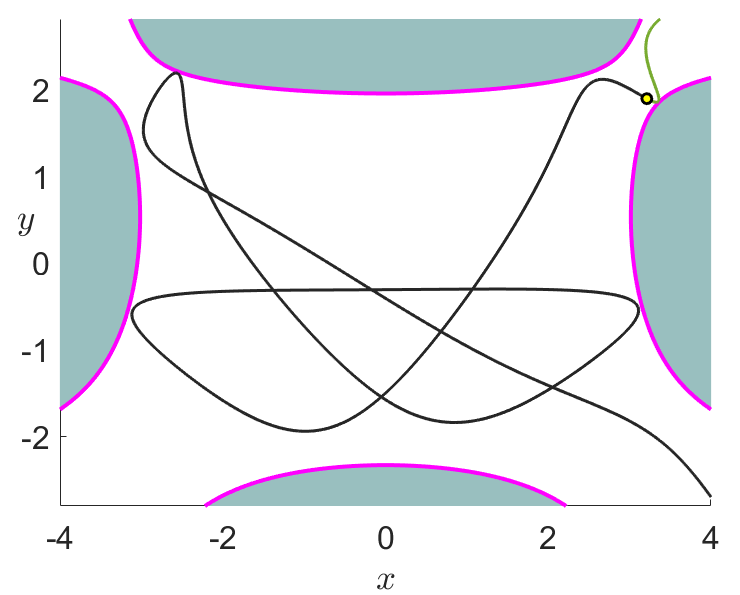}\\
	\caption{Phase space structures calculated on the Poincar\'e section described in Eq. \eqref{psos} located in the vicinity of the UPO of the upper-right index-1 saddle. We have similar structures in the vicinity of the UPO of the upper-left index-1 saddle because of the symmetry of the potential. We illustrate the formation of an heteroclinic connection, as the stretching parameter of the Caldera PES is varied, between a stable manifold (blue curve) of an UPO of the central region of the Caldera and the unstable manifold (red curve) of the UPO associated to the upper index-1 saddle. The first row corresponds to $\lambda = 0.8$, the second row is for the critical stretching value $\lambda = 0.778$, and for the third row we use $\lambda = 0.7$. In the second column, we capture the inhibition of dynamical matching by depicting the projection onto configuration space of the trajectory of an initial condition marked as a yellow dot. Forward and backward evolution of the trajectory are represented in black and green respectively. The magenta curve represents the energy boundary.}
	\label{fig_panel}
\end{figure}

For a two DoF system, the fixed energy surface is three dimensional. For energies above that of the upper saddles  an unstable periodic orbit exists in the energy surface. This is a consequence of the Lyapunov subcenter manifold theorem \cite{moser1976, weinstein1973, rabinowitz1982}. In a fixed energy surface, these periodic orbits have two dimensional stable and unstable manifolds. Trajectories move away from the periodic orbits along the direction of the unstable manifold in forward time. In the upper left panel of Fig. \ref{fig_panel} we show a segment of the unstable manifold of the upper right-hand saddle directed towards the interior of the caldera.

The region of the central minimum of the caldera may also contain unstable periodic orbits. The stable manifolds of these periodic orbits direct trajectories towards the central minimum. In the upper left panel of Fig. \ref{fig_panel} we show a segment of the stable manifold of an unstable periodic orbit in the region of the central minimum directed away from the central minimum.

If the stable manifold of a periodic orbit in the central minimum intersects the unstable manifolds of one of the  upper  saddles we have a mechanism for trajectories to enter the caldera and be directed towards the region of the central minimum. In dynamical systems terminology this is referred to as a heteroclinic connection. This would inhibit dynamical matching, as trajectories entering the caldera would exhibit (temporary) trapping in the region of the central minimum. If the heteroclinic connection breaks, as might occur if a parameter is varied, the mechanism for directing trajectories towards the regions of the central minimum no longer exists, and dynamical matching is possible.  Hence, a heteroclinic bifurcation is the critical phase space structure that inhibits or allows dynamical matching, which we now show.

In order to explore the formation of a heteroclinic intersection between any stable manifold coming from an UPO of the central region of the Caldera and the unstable manifold of the UPO of the upper-right index-1 saddle, as the stretching parameter of the Caldera PES is varied, we probe the phase space structures in the following Poincar\'e surface of section:
\begin{equation}
\mathcal{U}^{+}_{x,p_x} = \lbrace (x,y,p_x,p_y) \in \mathbb{R}^4 \;|\; y = 1.88409 \; ,\; p_y > 0 \;,\; E = 29 \rbrace
\label{psos}
\end{equation}

In the middle-left panel of Fig. \ref{fig_panel}, we observe that there is a critical value of stretching parameter ($\lambda=0.778$) for the formation of this heteroclinic connection. For values of the stretching parameter above the critical value there is no  heteroclinic connection between any stable manifold coming from an UPO of the central region of the Caldera and the unstable manifold of the UPO of the upper index-1 saddle (see the upper left   panel of Fig. \ref{fig_panel}). The non-existence of these  heteroclinic connections results in the phenomenon of dynamical matching. In this case, if we integrate an initial condition inside the region of the unstable manifold of UPO of the upper-right index-1 saddle forward and backward in time, we see in the upper right panel of  Fig. \ref{fig_panel} that the resulting trajectory comes from the region of the upper-right index-1 saddle and exits the caldera through the region of the opposite lower saddle without any interaction with the central area of the caldera.

Now, for values of the stretching parameter equal or above the critical value we have the formation of heteroclinic connections between  the stable manifold coming from an UPO of the central region of the Caldera and the unstable manifold of the UPO of the upper-right index-1 saddle, see (middle and lower left panels of Fig. \ref{fig_panel}. This heteroclinic connection destroys the dynamical matching mechanism because many trajectories become trapped inside the lobes between the two invariant manifolds. We can see this better if we choose an initial condition inside a lobe, as we illustrate in the middle and lower left panels of Fig. \ref{fig_panel}) and integrate it forward and backward. We observe that the resulting trajectory is temporarily trapped in the central area of the caldera before it exits from this area, see the middle and lower right panels of Fig. \ref{fig_panel}. 

\section{Summary and Outlook}
\label{CONCL}

In this paper we have shown that heteroclinic connections are the phase space mechanism that controls dynamical matching. While we have demonstrated this behavior for a two DoF caldera model, the notion of a heteroclinic trajectory is valid for dynamical systems with an arbitrary number of dimensions. Hence, it would be interesting to explore the formation of this phase space structure as a mechanism for dynamical matching in systems with three or more DoF.

\section*{Acknowledgements}
The authors acknowledge the support of EPSRC Grant No. EP/P021123/1 and Office of Naval Research Grant No. N00014-01-1-0769.

\bibliography{caldera2c}

\begin{thebibliography}{10}
\expandafter\ifx\csname url\endcsname\relax
  \def\url#1{\texttt{#1}}\fi
\expandafter\ifx\csname urlprefix\endcsname\relax\def\urlprefix{URL }\fi
\expandafter\ifx\csname href\endcsname\relax
  \def\href#1#2{#2} \def\path#1{#1}\fi

\bibitem{baldwin2003}
J.~Baldwin, Thermal rearrangements of vinylcyclopropanes to cyclopentenes,
  Chemical reviews 103~(4) (2003) 1197--1212.

\bibitem{gold1988}
Z.~Goldschmidt, B.~Crammer, Vinylcyclopropane rearrangements, Chem. Soc. Rev 17
  (1988) 229--267.

\bibitem{doubleday1997}
C.~Doubleday, K.~Bolton, W.~Hase, Direct dynamics study of the stereomutation
  of cyclopropane, Journal of the American Chemical Society 119~(22) (1997)
  5251--5252.

\bibitem{doubleday1999}
C.~Doubleday, M.~Nendel, K.~Houk, D.~Thweatt, M.~Page, Direct dynamics
  quasiclassical trajectory study of the stereochemistry of the
  vinylcyclopropane - cyclopentene rearrangement, Journal of the American
  Chemical Society 121~(19) (1999) 4720--4721.

\bibitem{doubleday2006}
C.~Doubleday, C.~Suhrada, K.~Houk, Dynamics of the degenerate rearrangement of
  bicyclo[3.1.0]hex-2-ene, Journal of the American Chemical Society 128~(1)
  (2006) 90--94.

\bibitem{reyes2002}
M.~Reyes, E.~Lobkovsky, B.~Carpenter, Interplay of orbital symmetry and
  nonstatistical dynamics in the thermal rearrangements of
  bicyclo[n.1.0]polyenes, J Am Chem Soc 124 (2002) 641--651.

\bibitem{carpenter1985}
B.~K. Carpenter, Trajectories through an intermediate at a fourfold branch
  point. implications for the stereochemistry of biradical reactions, Journal
  of the American Chemical Society 107~(20) (1985) 5730--5732.
\newblock \href {http://dx.doi.org/10.1021/ja00306a021}
  {\path{doi:10.1021/ja00306a021}}.

\bibitem{carpenter1995}
B.~K. Carpenter, Dynamic matching: The cause of inversion of configuration in
  the [1,3] sigmatropic migration?, Journal of the American Chemical Society
  117~(23) (1995) 6336--6344.
\newblock \href {http://dx.doi.org/10.1021/ja00128a024}
  {\path{doi:10.1021/ja00128a024}}.

\bibitem{collins2014}
P.~Collins, Z.~Kramer, B.~Carpenter, G.~Ezra, S.~Wiggins, Nonstatistical
  dynamics on the caldera, Journal of Chemical Physics 141~(034111).

\bibitem{katsanikas2018}
M.~Katsanikas, S.~Wiggins, Phase space structure and transport in a caldera
  potential energy surface, International Journal of Bifurcation and Chaos
  28~(13) (2018) 1830042.

\bibitem{katsanikas2019}
M.~Katsanikas, S.~Wiggins, Phase space analysis of the nonexistence of
  dynamical matching in a stretched caldera potential energy surface,
  International Journal of Bifurcation and Chaos 29~(04) (2019) 1950057.

\bibitem{mancho2013lagrangian}
A.~M. Mancho, S.~Wiggins, J.~Curbelo, C.~Mendoza, Lagrangian descriptors: A
  method for revealing phase space structures of general time dependent
  dynamical systems, Communications in Nonlinear Science and Numerical
  Simulation 18~(12) (2013) 3530--3557.
\newblock \href {http://dx.doi.org/10.1016/j.cnsns.2013.05.002}
  {\path{doi:10.1016/j.cnsns.2013.05.002}}.

\bibitem{lopesino2017}
C.~Lopesino, F.~Balibrea-Iniesta, V.~J. Garc\'ia-Garrido, S.~Wiggins, A.~M.
  Mancho, A theoretical framework for {L}agrangian descriptors, International
  Journal of Bifurcation and Chaos 27~(01) (2017) 1730001.
\newblock \href {http://dx.doi.org/10.1142/S0218127417300014}
  {\path{doi:10.1142/S0218127417300014}}.

\bibitem{naik2019a}
S.~Naik, V.~J. García-Garrido, S.~Wiggins, Finding {NHIM}: Identifying high
  dimensional phase space structures in reaction dynamics using {L}agrangian
  descriptors, Communications in Nonlinear Science and Numerical Simulation 79
  (2019) 104907.
\newblock \href {http://dx.doi.org/10.1016/j.cnsns.2019.104907}
  {\path{doi:10.1016/j.cnsns.2019.104907}}.

\bibitem{madrid2009}
J.~A.~J. Madrid, A.~M. Mancho, {Distinguished trajectories in time dependent
  vector fields}, Chaos 19 (2009) 013111.
\newblock \href {http://dx.doi.org/10.1063/1.3056050}
  {\path{doi:10.1063/1.3056050}}.

\bibitem{craven2015lagrangian}
G.~T. Craven, R.~Hernandez, Lagrangian descriptors of thermalized transition
  states on time-varying energy surfaces, Physical review letters 115~(14)
  (2015) 148301.
\newblock \href {http://dx.doi.org/10.1103/PhysRevLett.115.148301}
  {\path{doi:10.1103/PhysRevLett.115.148301}}.

\bibitem{craven2016deconstructing}
G.~T. Craven, R.~Hernandez, Deconstructing field-induced ketene isomerization
  through {L}agrangian descriptors, Physical Chemistry Chemical Physics 18~(5)
  (2016) 4008--4018.
\newblock \href {http://dx.doi.org/10.1039/c5cp06624g}
  {\path{doi:10.1039/c5cp06624g}}.

\bibitem{craven2017lagrangian}
G.~T. Craven, A.~Junginger, R.~Hernandez, Lagrangian descriptors of driven
  chemical reaction manifolds, Physical Review E 96~(2) (2017) 022222.
\newblock \href {http://dx.doi.org/10.1103/PhysRevE.96.022222}
  {\path{doi:10.1103/PhysRevE.96.022222}}.

\bibitem{junginger2017chemical}
A.~Junginger, L.~Duvenbeck, M.~Feldmaier, J.~Main, G.~Wunner, R.~Hernandez,
  Chemical dynamics between wells across a time-dependent barrier:
  Self-similarity in the {L}agrangian descriptor and reactive basins, The
  Journal of chemical physics 147~(6) (2017) 064101.
\newblock \href {http://dx.doi.org/10.1063/1.4997379}
  {\path{doi:10.1063/1.4997379}}.

\bibitem{naik2019b}
S.~Naik, S.~Wiggins, Finding normally hyperbolic invariant manifolds in two and
  three degrees of freedom with {H}{\'e}non-{H}eiles type potential, Phys. Rev.
  E 100 (2019) 022204.
\newblock \href {http://dx.doi.org/10.1103/PhysRevE.100.022204}
  {\path{doi:10.1103/PhysRevE.100.022204}}.

\bibitem{GG2019}
V.~J. Garc\'{i}a-Garrido, S.~Naik, S.~Wiggins, Tilting and squeezing: {P}hase
  space geometry of {H}amiltonian saddle-node bifurcation and its influence on
  chemical reaction dynamics, arXiv preprint:1907.03322 ({\it Under Review}).

\bibitem{KGW2019}
M.~Katsanikas, V.~J. Garc\'{i}a-Garrido, S.~Wiggins, Detection of {D}ynamical
  {M}atching in a {C}aldera {H}amiltonian {S}ystem using {L}agrangian
  {D}escriptors, arXiv preprint: XXXX ({\it Under Review}).

\bibitem{moser1976}
J.~Moser, Periodic orbits near an equilibrium and a theorem by alan weinstein,
  Communications on Pure and Applied Mathematics 29~(6) (1976) 727--747.

\bibitem{weinstein1973}
A.~Weinstein, Normal modes for nonlinear hamiltonian systems, Inventiones
  mathematicae 20~(1) (1973) 47--57.

\bibitem{rabinowitz1982}
P.~Rabinowitz, Periodic solutions of hamiltonian systems: a survey, SIAM J.
  Math. Anal. 13~(3) (1982) 343--352.

\end{thebibliography}

\end{document}